\documentclass{article} \usepackage{amsmath} \usepackage{amssymb}
\usepackage{epsfig} \usepackage{verbatim} 
\newcommand{\cR}{\mathcal{R}} \newcommand{\cC}{\mathcal{C}}
\newcommand{\cD}{\mathcal{D}} \newcommand{\tcD}{\tilde{\mathcal{D}}}
\newcommand{\tphi}{\tilde \phi} 
 \newcommand{\mean}[1]{\langle #1 \rangle}
\newcommand{\tP}{\tilde{P}}

\newcommand{\tF}{\tilde{F}}

\newcommand{\vel}{\nu}

\newcommand{\bra}{\langle}
\newcommand{\ket}{\rangle}
\newcommand{\var}{var}
\newcommand{\Cov}{\textit{Cov}}

\newcommand{\ab}{(a)}
\newcommand{\bb}{(b)}

\newcommand{\jib}{(j_i)}
\newcommand{\jab}{(j_1)}
\newcommand{\jbb}{(j_2)}
\newcommand{\jcb}{(j_3)}
\newcommand{\jdb}{(j_4)}

\font\openface=msbm10 at10pt
\def\Minkowski {{\hbox{\openface M}}}
  \def\Reals {{\hbox{\openface R}}}

\def\tilde{\widetilde}          % define tilde to always be the ``widetilde''
\def\SetOf#1#2{\left\{ #1  \,|\, #2 \right\} }

\def\sqr#1#2{\vcenter{
  \hrule height.#2pt
  \hbox{\vrule width.#2pt height#1pt
        \kern#1pt
        \vrule width.#2pt}
  \hrule height.#2pt}}

\def\dal{\mathop{\,\sqr{7}{5}\,}}

\begin{document}

\title {Discreteness and the transmission of light from distant sources}

\author{Fay Dowker\footnote{Blackett Laboratory, Imperial
 College, London SW7 2AZ, UK.}, Joe Henson\footnote{Perimeter
 Institute, 31 Caroline St. N., Waterloo, N2L 2Y5, Canada} and Rafael
 D. Sorkin\footnote{Perimeter Institute, 31 Caroline St. N., Waterloo,
 N2L 2Y5, Canada.} }
\maketitle

\begin{abstract}
We model the classical transmission of a massless scalar field
from a source to a detector on a background causal set.  The
predictions do not differ significantly from those of the continuum.
Thus, introducing an intrinsic inexactitude to lengths and durations -- or more specifically, replacing the Lorentzian manifold with an underlying discrete structure -- need not disrupt the usual dynamics of propagation.
\end{abstract}
\vskip 1cm

%=====================================================

\section{Introduction}

Despite the variety of approaches the the problem of quantum gravity, it is
 hoped by some that agreement can be forged on suitably generic consequences of the
as yet unknown theory.  This would offer hope of deriving generic predictions of quantum gravity in a more or less heuristic fashion.  However, deriving such predictions often turns out to demand a greater level of specificity, whereby disagreements return.  For instance, a majority of workers would probably agree
that the differentiable manifold structure of spacetime will break
down near the Planck scale to be replaced by something of a more
discrete, ``quantised'' or foamy nature.  When potentially
observable consequences of these general ideas are sought,
 however, the consensus evaporates.
% In particular, some
% have argued that there is enough agreement to roughly formulate
% generic predictions motivated by quantum gravity; when a model is
% offered on which to base the predictions, however, controversy often
% returns.
For example, disagreement arises over whether the expected
break-down of General Relativity at Planck scales would give rise to modified
dispersion relations or other Lorentz symmetry violating phenomena.
Spacetime discreteness is often cited as
motivation to consider Lorentz symmetry violation (see \textit{e.g.}
\cite{AmelinoCamelia:2008qg}), perhaps due to consideration of altered
dispersion in simple lattice models.  There are also
quantum gravity inspired models that draw the same conclusion
\cite{Gambini:1998it,Alfaro:1999wd}.
%these papers are in the context
%of LQG and make no completely generic claim.
Meanwhile, there are models of discrete spacetime
which respect Lorentz invariance \cite{Bombelli:1987aa,
Dowker:2003hb,Bombelli:2006nm} and so do not result in
Lorentz symmetry violating phenomenology. Another example
of controversy is the arguments in references
\cite{Amelino-Camelia:2004yd, Coule:2003td, Ng:2003ag} which dispute the
generic claim of reference \cite{Lieu:2003ee} that any quantum
gravitational ``fuzzing'' of the metric would disturb
the coherence of light from distant sources by an amount greater than
that allowed by the observations.

The lesson is that it is important to test
expectations of phenomenological effects of quantum gravity
against different concrete models in order to determine if
they are generic.
One might indeed be tempted to claim that any ``fuzzing''
of spacetime properties -- be it quantum uncertainty in time and
distance measurements, lack of definition of short distances due to
discreteness, or from some other source -- would naturally lead to some
loss of coherence of light from distant sources (for a model inspired
by similar considerations, see \cite{Christiansen:2005yg}).  One
purpose of this paper is to test this expectation in a
definite model.  More specifically, the model examines this question in the discrete context
of causal set theory.  We present a model in which spacetime metric
relations do indeed have an approximate character that breaks down at
around the Planck scale, but in which no significant loss of phase
information results.

One crucial point is that, whereas other models introduce
Lorentz violating assumptions, the causal set model allows for Lorentz
invariance in the approximating continuum.  That the transmission of
light in this context must be consistent with Lorentz symmetry follows
already from the results of
\cite{Dowker:2003hb,Bombelli:2006nm}, and in agreement with
this conclusion, the model
we study below will exhibit the usual, strict proportionality of
frequency to wavelength.
However, we know of no equally general reasons that would ensure the
coherence of light from distant sources; and so one might wonder
whether the underlying discreteness would necessarily disrupt the
coherence of propagated waves, or wrinkle the wave fronts or alter
the wavelength,  and hence the frequency of the signal.

A second purpose of this paper is to learn something about other
possible effects of an underlying atomicity that one might expect to
encounter in analogy with more familiar examples like the propagation
of light through air or other material media.  Assuming this analogy
is valid, effects like scattering and extinction will be present at
some level, and the question becomes whether one can expect them to
rise to the level of observability with current technology.

%% In a model whose discreteness also broke Lorentz invariance, further
%% effects might ensue; the velocity of light could become frequency
%% dependent or could be subject to fluctuations.
%%
%% Whether such effects occur at all will of course depend on the details
%% of one's model.

In this paper we formulate a rudimentary model\footnote{Some features
of the model were previously reported in the proceedings of a conference \cite{Sorkin:2009bp}.} with whose aid one can
study some of these putative effects, assuming that the microscopic
structure which replaces the continuum at the Planck scale is the
causal set.  The specific process we will analyze is the transmission
of a signal from a source to a detector by a (classical) massless
scalar field in Minkowski spacetime $\Minkowski^4$.  Ideally, one
would analyze this process with respect to a complete theory of
electromagnetism formulated within the context of a complete theory of
quantum gravity.  In practise, however, one can hope that a simple
model incorporating the elements of discreteness and wave-propagation
can illuminate the range of possibilities to be expected.  Here we
study the simplest such model we could devise, based on a classical
scalar field $\phi$ of zero mass.  We do not formulate an independent
dynamics for $\phi$.  Rather we describe its transmission from source
to detector in terms of the discrete analog for the causal set of the
retarded Green's function (a delta function on the future light cone)
that yields the Lienard-Wiechert potentials in the continuum.  In the
future, we hope to compare the predictions of this model with those of
a more complete theory of wave propagation in a causal set (for progress
in this direction, see \cite{Henson:2006kf,Sorkin:2007qi, Johnston:2009fr,Benincasa:2010ac}).

\section{A continuum model}

\begin{figure}[ht]
\centering \resizebox{4.4in}{2.4in}{\includegraphics{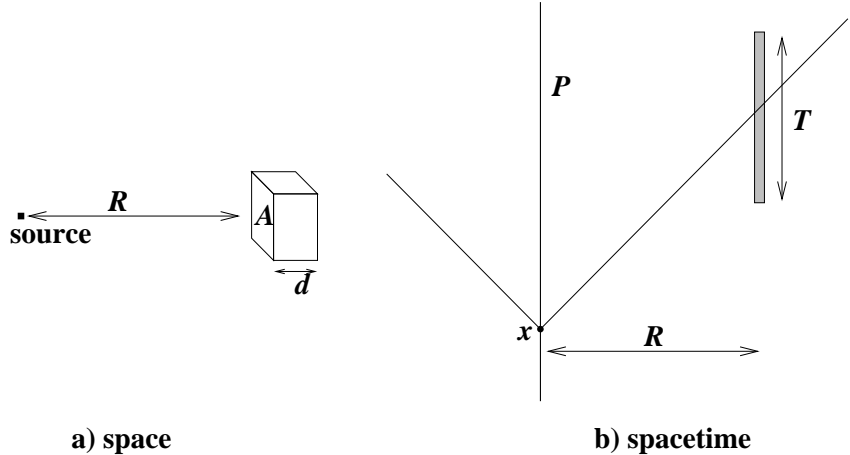}}
\caption{\small{The Source and Detector.  In (a) the spatial layout is
 shown, while (b) is a spacetime diagram (with only one spatial
 dimension represented) showing the source's worldline $P$, the
 light-cone of a point $x$ on that worldline, and the detector region
 in grey.
%% Normal conventions for spacetime axes are used.
}\label{detector}}
\end{figure}

As illustrated in figure \ref{detector}, we will imagine an idealised
setup consisting --- in a continuum description --- of an oscillating
point source of scalar charge $q$, together with a detector of
rectangular shape, facing the source and at rest with respect to it in
an ambient flat spacetime\footnote
{For a slightly more realistic model (at least for the comparison to propagation of electromagnetic radiation) the source should be neutral.  To that end we will later be supplementing this point-charge with a
static companion of opposite sign; this cancels the constant, ``DC'' component of
the far field but leaves the AC component unaffected.  We could also include more worldlines to
build up an extended source, say in the dipole approximation.  On the
other hand, we could (since scalar charges need not be conserved)
simplify our setup even further by keeping the charge at rest but
letting its strength $q$ vary with time.  This would take us farther
from the astrophysically relevant, electromagnetic case, however.}
.  In a system of Cartesian coordinates in which the detector is at rest,
let the source oscillate about the origin, tracing out the trajectory
\begin{equation}
    \label{trajectory} x^1 = a \cos{\omega x^0} \,, \ x^2=x^3=0 \ ,
\end{equation}
where $a$ is a negligible fraction of the distance $R$  to the detector.
(On average, the source is thus at rest with respect to the detector,
another simplification which one could easily relax.)  Assuming that
the detector has a resolution time of $T$, a two-dimensional area $A$,
and a thickness $d$, this gives us in spacetime a {\it detection
region} $\cD$ of volume $TAd$, which we can take to be the rectangular
domain $[R,R+T]\times[R,R+d]\times[0,\sqrt{A}]\times[0,\sqrt{A}]$.
(In this paper we adopt Planckian units in which $c=\hbar=8{\pi}G=1$.)
%
%  $R = L + a \cos{\omega t}$.
%
Let us assume further that $R$ greatly exceeds
$\lambda\sim\omega^{-1}$, the wavelength of the emitted radiation,
which in turn is much greater than any detector dimension; and also
that the detector's response rate is rapid compared to the period of
the source oscillations.  We have then $$ d, \, \sqrt{A} \ll \lambda
\ll R ,\ \ T \ll \lambda ,\ \ {\rm and}\ a\ll R \ .  $$ Up to small
corrections, the spatial distance between emitter and receiver is thus
$R$.
We remark that, in order to observe coherence effects, the detector
actually must be operational for a much longer time than $T$, namely a
time comparable to $\omega^{-1}$.  Finally, let us take the ``output''
$F$ of our detector
%% at time $t$
to be simply the integral of the field, $\phi$, over the detection region $\cD$.
%
%% over the portion of the ``detector world tube'' occupying the
%% time-interval $[t-T, t]$
%
\footnote
{This model could be regarded as fairly realistic for radio wave
reception.  For optical frequencies, the time-resolution of realistic
detectors is not better than the period of the wave, and the received
signal is usually compared to another oscillator in order to diminish
the effective value of $T$.  In such cases our parameter $T$ must be
understood as an effective resolution-time characterizing the
detector.  }.
%%  If we were to build a model of detection in this vein, the detector
%%  region would be larger, and so this only would have made the results in
%%  the following discrete model closer to those in the continuum.
%  JH:Something like this necessary?
%  FD:Not sure its necessary but it heads off a possible objection so it's useful
%% [[RDS  1. reworded. 2. See my query on this]]

Before proceeding to our causal set model of the same setup, let us
first review the treatment of this situation in the continuum.  The
retarded Green's function for a massless scalar field in
$(3+1)$-dimensions is a delta-function on the forward light-cone:
\begin{align}
  \label{Green} G(y,x) &= \begin{cases} \frac{1}{2 \pi} \delta(|y -
  x|^2) & \text{if $y$ is in the causal future of $x$} \\ 0 &
  \text{otherwise} \\ \end{cases} \\ &= \frac{1}{4 \pi r} \delta (y^0
  - x^0 - r),
\end{align}
where $r$ is the spatial distance from $x$ to $y$.  In terms of $G$,
the field produced by our source is given by
\begin{gather}
  \phi(y) = \int_P \, G(y,x(s)) \, q \, ds \ ,
\end{gather}
%%  \phi(y) = q \int_P \, ds \, G(x(s), y) \ ,
where $P$ is the worldline of the source, $q$ is its charge, and $s$
is proper time along $P$.\footnote
{We have normalized $G$ so that $\dal G = -\delta$, where our metric
signature is $(- + + +)$.
%%  and the coupling between $\phi$ and the source is that corresponding to
%%  the action functional $-\int q(s)\phi(x(s))ds = -\int\phi(x)j(x)d^4x$,
%%  $s$ being proper time along the source worldline.  (Varying the last
%%  integral yields the field equation $\dal\phi=-j$.)
 Notice that only an infinitesimal portion of $P$ contributes to
 $\phi(y)$.}
According to the ansatz we have made, the output of the detector is
proportional to the integral $\int_{\cD} \, d^4y \, \phi(y)$, and thus
can be expressed directly in terms of the source and the Green's
function:
%% of $\phi$ over the region $\cD$:
\begin{gather}
\label{e:Fdoubleint}
  F = q \int_P \, ds \int_{\cD} \, d^4y \, G(y,x(s))
\end{gather}
% changed ``\propto'' to ``= q'' here.
The two integrals can be done in either order.  Integrating first
with respect to $s$ yields for $\phi(y)$
\begin{equation}
      \frac{q}{4 \pi R} \; \frac{ds/dt }{ 1 - dx/dt} \ , \label{phi}
\end{equation}
where $ds/dt$ and $dx/dt$ are to be evaluated at the retarded time
$t-R$ corresponding to the detection time $t$ of interest.  (Strictly
speaking, this expression is correct only at a single point within the
detector close to $(R,0,0)$, but we can ignore this caveat, since we
have assumed the detector to be small compared to $R$ and $T$ small
compared to the period of oscillation.)\footnote
{The fact that different points in the detector correspond to
different values of $s$ could not have been ignored if we had
integrated first over $y$, however.}
In (\ref{phi}), the oscillating signal resides in the time-dependent
factor multiplying $1/R$, which can be understood as just the Doppler
effect in disguise.  The $y$-integration is now trivial and furnishes
for the detector output
\begin{equation}
\label{e:continuumoutput}
        F \approx \sqrt{\frac{1+\vel}{1-\vel}} \; \frac{q}{4 \pi R} \;
        TAd \ ,
\end{equation}
where $\vel=dx/dt$ is the component of the source velocity toward the
detector at the relevant retarded time.
 In a slow velocity
approximation, $\sqrt{1+\vel/1-\vel}$ reduces to $1+\vel$, and the
varying, ``AC''
part of the signal is a multiple of $\sin(\omega t)/4\pi R$.

In order explicitly to subtract off the DC part of the signal
and create a more
realistic model of electromagnetic radiation, we can add a static
negative charge, $-q$, at the origin of spatial coordinates. Then,
renaming $F$ above as $F_+$ and calling the detector response to the
negative charge, $F_-$, we have the total detector response
\begin{equation}
 F_{\textrm{total}} := F_+ + F_- = \sqrt{\frac{1+\vel}{1-\vel}}\; \frac{q}{4\pi R} T A \,d -
\frac{q}{4\pi R} T A \,d
\end{equation}
which, for low velocities is
\begin{align}
F_{\textrm{total}} & =\vel \frac{q}{4\pi R} T A \,d\\
{}& =  \frac{q}{4\pi R} T A\, d\, a\, \omega \sin(\omega t)\,.
\end{align}
The detector response is coherent in space as well as in time,
so our model captures the coherence that spacetime uncertainty might
be expected to disrupt.

Note that the $1/R$ dependence of the signal agrees with the
usual falloff of far field electromagnetic radiation from a dipole source.
Here the signal is proportional to the velocity of the source however,
whereas the signal in the electromagnetic case is proportional to the
acceleration. We note that we could make the model more
realistic by pretending that the ``field'' $\phi(y)$ is the electrostatic
potential $A_0$ of an electromagnetic field. The relevant spatial gradient of
$A_0 = \phi(y)$ would arise from differential time delay (detectors at
different spatial positions see the source at different times along
its world line) and could give rise
to an electric field proportional to the acceleration of the
source.

%% Because the position of the source is effectively constant over the
%% relevant period, the integral over $s$ can be rewritten:
%% \begin{equation}
%%   \phi(y) \approx q \int \, dx^0 \, G(x, y).
%% \end{equation}
%% %% Let as designate as $(0,1,0,0)$ the direction from the source to the
%% %% detector.
%% The intersection of the forward light-cone of any source
%% point with the detector region, when it does intersect, is well
%% approximated by a plane.  So
%% \begin{align}
%% \phi(y) &\approx \frac{q}{4 \pi R} \int \, dx^0 \, \delta \bigl(
%% (y^0-x^0)-(y^1-x^1) \bigr) \\ &\approx \frac{q}{4 \pi R},
%% \end{align}
%% and so we see that $\phi(y)$ is approximately constant over the
%% detector region, giving

%% %[ JH: Changed so that we do not require $ d \ll T$ ]
%% The detector
%% output tracks the distance from the source, so as the source
%% oscillates around its mean position, the output varies with the same
%% frequency.

\section{The discrete model}

Now let us consider a causal set model for the same situation.  Before
turning to the model \textit{per se} let us recall some kinematical
results from causal set theory.  (For a more detailed introduction see
\cite{Bombelli:1987aa, Sorkin:1990bh,Henson:2006kf}.)  A causal set $\cC$, or causet
for short, is a locally finite, partially ordered set.  When a causal
set has an approximation by a continuum spacetime, the order-relation
of $\cC$ corresponds to the causal order of spacetime, while the
number of elements in a subset of $\cC$ equals (up to fluctuations)
the volume of the corresponding spacetime region in fundamental units.
It is conjectured that this information is all that is needed to
recover the metric, differential structure, dimension and topology of
the approximating Lorentzian manifold (at scales large compared to the
fundamental scale).

With a nonzero discreteness scale, the correspondence of any
Lorentzian manifold to a causet can only be approximate, and it seems
to be necessary to define it stochastically.  Although the causet is
taken to be fundamental and the manifold is merely an effective
description of it on large scales, some means must be found of
determining whether a given spacetime approximates to a certain causal
set.  To this end we consider a method called \textit{sprinkling},
which produces a random causet from a given Lorentzian manifold by
means of a Poisson process \cite{Stoyan:1995}.  One samples, at
random, points from the manifold with a density of $1$ in some
fundamental units.  (One might term these fundamental units ``causet
units''.  They are the analog of lattice units in lattice gauge
theory, and one would naturally expect them to agree approximately
with Planck units as defined above; for the purposes of this paper
we will take the agreement to be exact.  To determine the true factor
relating the two sets of units would be to determine the fundamental
discreteness scale and would be a key step in the development of the
theory.)  A causet can then be constructed using these ``sprinkled''
points as elements, the order-relation being that induced by the
causal order of the Lorentzian manifold.  A manifold is defined to
approximate a causal set when the causal set is a typical result of
the sprinkling process on that manifold.

Often, instead of finding an approximating spacetime for a given
causal set, we are given the approximating spacetime, and wish to find
a causal set to which it approximates.  In that case, we can reverse
the argument and construct possible underlying causal sets using the
sprinkling process.  If, in the sprinkling process, a property of the
resulting causal set holds with probability 1, we say that the
property holds for a typical sprinkling.  Similarly, properties that
hold with probabilities sufficiently near to 1 can be considered
typical.  Some care is needed to properly define this notion in all
cases.  However, without fully entering into this discussion, if the
only quantity of interest is a function $x(\cC)$ of the causal set,
the value for a typical sprinkling will be only a small number of
standard deviations away from the mean over sprinklings $\mean{x}$.
One can imagine cases in which $x$ is
an ``ensemble average'' of many
variables pertaining to different regions.

In a Poisson process, the probability to sprinkle $n$ elements into a
spacetime region of volume $V$ is given (for any measurable subset) by
the Poisson distribution,
\begin{equation}
               P(n)= \frac{V^n e^{-V}}{n!} \ .
\end{equation}
Because a Poisson process depends only on the spacetime
volume-element, it is Lorentz invariant.  This invariance is exact for
$d$-dimensional Minkowski spacetime, where the theorems on the
existence and uniqueness of the Poisson process readily establish its
invariance under all volume preserving linear maps, and in particular
under arbitrary Poincar{\'e} transformations (see {\it e.g.} reference
\cite{Stoyan:1995}).
%% with the distribution $P(n)$ for all measurable subsets of $\Reals^d$
%% and its invariance under all volume preserving linear maps establish
%% Lorentz invariance rigorously.
For a proof of invariance that holds even for individual
realizations, see \cite{Bombelli:2006nm}.  Thus, if the causal set is
well-approximated by Minkowski space at all, then the discreteness
does nothing to pick out a preferred frame in it. In a curved
spacetime, Lorentz invariance is to be understood in the same, local
and approximate sense that holds in General Relativity.

For future reference we define the terms ``interval'', ``chain'',
``path'' and ``link''.  Given elements $p$ and $q$ in $\cC$, the {\it
order-interval} between them is $I(p,q)=\SetOf{r\in \cC}{p\prec r\prec
q}$, where $\prec$ is the fundamental precedence relation defining
$\cC$.  A {\it chain} is a linearly ordered subset of $\cC$ and a {\it
path} is a saturated chain, i.e. one which is maximal within the
order-interval between any two of its elements.  A \textit{link} is a
causal relation $p \prec q$ between two elements, $p$ and $q$, that is
not implied via transitivity by other relations, i.e. there exists no
third element $r$ such that $p \prec r \prec q$.

We now turn to the causet description of the source-detector system.
We will describe the discrete counterparts of the spacetime, the
source and detector, and the model of propagation used above, and
determine the detector output in this model.  In place of
4-dimensional Minkowski space $\Minkowski^4$, we will have a typical
sprinkling of $\Minkowski^4$, and in place of the detector region, the
subset, $\tcD$, comprising all the elements that were sprinkled into
the region of spacetime $\cD$.  The field $\phi$ will be a function
from the elements of the causet to $\Reals$.

We will ignore, for now, the static
negative charge source and study only the discrete analogue of the
oscillating positive charge. In analogy
with the continuum model of the source as a point-charge,
we may in the causet identify the source with a {\it path} $\tP$ that
approximately follows the spacetime worldline $P$ of eqn.
(\ref{trajectory}).  Such a path will be ``locally geodesic'' in the
sense of approximating a longest chain between any two sufficiently
nearby elements of $\tP$.  We will say that an element of $\tP$ is
associated to a point $x$ of $P$ if it lies at a larger spacelike
distance from $x$ than any other point in $P$.  The sprinkled points
making up such a $\tP$ will be distributed along the corresponding
continuum curve with an approximate spacing in proper time along the
curve that is near to one Planck unit \cite{Brightwell:1990ha,Rideout:2008rk}.  (The ratio of
chain-length in the causet to proper length in the continuum does not
seem to be known exactly in 4-dimensions.  Simulations suggest a value
somewhat greater than 1.2.  This gives the average number of elements
per length of $P$ and this, along with the independence of the
sprinkling process in different regions, turns out to be all that is
needed for our calculations.) Such a locally geodesic path will exist,
and the spatial distance in the sprinkling from the curve $P$ to the
points corresponding to elements of $\tP$ will typically be a small
number in Planck units.

Both of these definitions make use of regions in the spacetime
$\Minkowski^4$ that we are to sprinkle, rather than being defined
``intrinsically'' by relations and/or fields on the causal set itself.
Defining them by such means would be more satisfying physically but
would complicate matters considerably, and if it led only to a change
in the position of the two regions on or around the order of the
Planck scale, that would not affect the results, as will become
obvious during the calculation.

%% The source is now modelled as the subset, $\tP$, of the causet which
%% lies in a tube of unit Planck cross section -- in the rest frame of the
%% detector -- centred on the continuum worldline of the source.
%% The expected number of elements in a segment
%% of the tube of proper length $\tau$ is $\tau$ and the Poisson
%% distribution means that the larger $\tau$, the more accurately the
%% expected value approximates the actual value.
%% The fluctuations grow as $\sqrt{\tau}$ but the relative size of
%% fluctuations to the mean decreases as $1/\sqrt{\tau}$.

In the continuum, the relevant dynamics of the field were completely
described by a retarded Green's function $G(x,y)$.  To define an
analogous ``propagator'' in the causal set, we must discover a
discrete replacement for $G(x,y)$ that approximates it well on large
scales, but relies only on the structure of the causal set, without
appealing to any extra information from the continuum.  This idea has
already been explored in \cite{Daughton:1993}, \cite{Salgado:2008}
(which are concerned primarily with the 2D situation) and \cite{Johnston:2008za}.

The task of finding a discrete propagator is made easier by the simple
form of equation (\ref{Green}), as we now describe, using the link
concept defined above.
%the continuum Green's function in \Minkowski^4
In a sprinkling of $\Minkowski^4$, the future links from any given
sprinkled point $e_i$ are unlikely to stretch over long proper times.
Nevertheless, thanks to the Lorentz invariance of the Poisson process,
links from $e_i$ will connect it with an infinite number of other
sprinkled points (``future nearest neighbors'') spaced out along, and
just inside of, the future light-cone.  This assertion can be proved
as follows.

The mean number of links joining an element at point
$x\in\Minkowski^4$ with elements in a region $\cR$ contained within
its future is
\begin{equation}
    \label{meanlinks} \mean{n_x} = \int_{\cR} \, d^4 y \,
    e^{-|I(x,y)|} \ ,
\end{equation}
where $|I(x,y)|$ is the spacetime volume of $I(x,y)=J^+(x) \cap
J^-1(y)$, the causal order-interval (or ``Alexandrov set'') between
$x$ and $y$.  This expression can be understood as the sum over
infinitesimal volumes $d^4y$ in $\cR$ of the probability that an
element is sprinkled into $d^4y$
%% at $y$ ({\it i.e.} $d^4 y$)
times the probability that this element is linked to the element at
$x$, {\it i.e.} the probability that no point was sprinkled into the
interval $I(x,y)$.  For a Poisson process, the former probability is
$d^4y$ itself and the latter is $e^{-|I(x,y)|}$, from which
(\ref{meanlinks}) follows.  Now in four dimensions,
\begin{equation}
   \label{interval} |I(x,y)| = \frac{\pi}{24} \tau^4(x,y) \ ,
\end{equation}
where $\tau(x,y)$ is the proper time between $x$ and $y$.  From this
one sees that an element that was sprinkled a large proper time from
$x$ is highly unlikely to be linked to it; but by the same token,
elements hugging the future light-cone of $x$ which are close to it in
proper time are quite likely to be linked to it.

Indeed, consider an element $e_0$ sprinkled at the point
$x\in\Minkowski^4$.  Choose coordinates $\xi, \eta, \theta, \varphi$
($\eta\ge0$), in which $x$ is at the origin, $\xi=0$, and the metric
is
\begin{equation}
       ds^2 = - d\xi^2 + \xi^2 d\eta^2 + \xi^2 \sinh^2\eta d\Omega_2^2
\end{equation}
so that $\xi$ is geodesic proper time from the origin.  The expected
number of future links from $e_0$ terminating in the region between
the light-cone of $x$ and the hyperboloid of points at a fixed proper
time $\tau$ from $x$ is given by (\ref{meanlinks}) with $\cR$ defined
by $0\le \xi \le \tau$, that is by
\begin{equation}
    4\pi \left[ \int_{0}^{\tau} d\xi \, \xi^3 e^{-\frac{\pi}{24}\xi^4}
    \right] \int_0^\infty d\eta \sinh\eta \ .
\end{equation}
Since the integral over $\eta$ diverges, the expected number of links
is infinite, no matter how small $\tau$ is.

%% [[RDS  <hapa> continue editing from here!]] %%%

We introduce the causet function
\begin{equation}
\label{e:ldef}
L(e',e)=
\begin{cases} \kappa & \text{whenever $e \prec e'$ and $\{e,e'\}$ is a link,}
\\
0 &\text{otherwise,}
\end{cases}
\end{equation}
where $e,e' \in \cC$ are causet elements and $\kappa$ is a normalising
constant of order 1, to be decided later.  In the limit of infinitely dense
sprinkling,
%
%% at which a sprinkled element is identified with every point of the
%% Lorentzian manifold,
%% RDS: not every point, but just a dense set of them
%
this function becomes a $\delta$-function on the forward light-cone of
$e$.  This means that $L(e',e)$ is a Lorentz invariant
discretisation of the continuum Green's function, and can be used to
define the propagation of the scalar field on the causet.

\subsection{Mean Detector Output}

We will calculate the average and variance of the detector output over
sprinklings of $\Minkowski^4$.  Replacing $G(x,y)$ with $L(e',e)$, the
calculation of the detector output is very similar to that in the
continuum, although now it is a random variable. The double integral
of eqn. (\ref{e:Fdoubleint}) for the output of the detector $\tilde F$
becomes a double sum over all elements in the source and detector:
\begin{equation}
  \tilde F=q \sum_{e \in \tP} \sum_{e' \in \tcD} L(e',e),
\end{equation}
%%%%%%%%%%%%%%%%%%%%%%%
\begin{comment}
some other formulations I took out:
\begin{gather}
  \tilde F=\sum_{e' \in \tcD} \tphi(e') \\ \tphi(e')=q \sum_{e \in
  \tP} L(e',e)
\end{gather}
In other words $\tilde F$ is proportional to the number of links from
elements in the source to elements in the detector.  We can reverse
the order of summation, defining
\begin{gather}
  \tF=\sum_{e \in \tP} \tF(e) \\ \tF(e)=q \sum_{e' \in \tcD} L(e',e').
\end{gather}
\end{comment}
%%%%%%%%%%%%%%%%%%%%%%%%
In other words $\tF$ is proportional to the number of links from
elements in the source region to elements in the detector region.  The
causal set is randomly generated by the sprinkling process in
$\Minkowski^4$, with the source and detector regions as described
above.  Using the definition of $L(e',e)$ in (\ref{e:ldef}), $\tF$ can
be rewritten as
\begin{equation}
\label{e:Frandvar}
\tF = q \kappa \int_{P} \int_{\cD \, x(s) \prec y} \sigma(ds) \chi(d^4y) \zeta(x(s),y) ,
\end{equation}
where $s$ and $x(s)$ are as in the continuum calculation, and
$\sigma(ds)$, $\chi(d^4x)$ $\zeta(x,y)$ are random variables in the
sprinkling process: the variable $\sigma(ds)=1$ if any point of $P$
with proper time in $ds$ has a sprinkled element of $\tP$ associated
to it and is 0 otherwise, $\chi(d^4y)=1$ if the volume element $d^4y$
contains a sprinkled point and 0 otherwise, and $\zeta(x,y)=1$ if the
causal interval between $x$ and $y$ (not including the points $x$ and
$y$) is empty of sprinkled points and 0 otherwise.  From this, we can
find the mean value:
\begin{align}
\mean{\tF} &= q \kappa \int_{P} \int_{\cD \, x(s) \prec y} \bra \sigma(ds) \chi(d^4y) \zeta(x(s),y) \ket , \\
    &= q \kappa \int_{P} \int_{\cD \, x(s) \prec y} \mean{\sigma(ds)}
    \mean{\chi(d^4y)} \mean{\zeta(x(s),y)} , \\ &= q \kappa \int_{P}
    \int_{\cD \, x(s) \prec y} ds \, d^4y \, e^{-|I(x(s),y)|} .
\label{e:meanF}
\end{align}
As explained above, the average number of points in $\tP$ associated
to $P$ per unit length, $\mean{\sigma(ds)}/ds$, is some number of
order unity, which we absorb into $\kappa$.  The calculation is
unaffected by the Planck scale deviation of the spatial positions of
elements in $\tP$ from $P$.  The second line above says that the three
random variables are always uncorrelated: firstly $x(s)$ and $y$ range
over disjoint regions, so $\sigma(ds)$ and $\chi(d^4y)$ are
uncorrelated, and secondly points sprinkled at $x(s)$ and $y$ have no
effect on $\zeta(x(s),y)$.  Let us first calculate
\begin{equation}
\label{e:meanFs}
\mean{\tphi(y)} =  q \kappa \int_{P \, x(s) \prec y}  \, ds \, e^{-|I(x(s),y)|} ,
\end{equation}
for $y \in D$.

(Despite the name this quantity is different to the discrete field,
which is a function of the causal set elements).  It is convenient to
define $r=y^1-x^1(s)$ and $t=y^0 - x^0(s)$, and use the co-ordinates
$u= t - r$, $v= t + r$ to calculate the integral:
\begin{equation}
\label{e:meanphi2}
\mean{\tphi(y)} \approx
  q \kappa \int_{u(s)=\infty}^{u(s)=0} \, ds \,
  e^{-\frac{\pi}{24}(uv)^2 } ,
\end{equation}
where we have used (\ref{interval}) and used $\sqrt{A} \ll R$ to
approximate the proper time between $x(s)$ and $y$.  Let us define
$R_y=r|_{u= 0}$.  Using this,
\begin{equation}
v=\alpha u + 2 R_y \, ,
\end{equation}
where
\begin{equation}
\alpha=\frac{1+\vel}{1-\vel}\,.
\end{equation}
We also have
\begin{equation}
ds=\sqrt{1-\vel^2} \, dx^0 =- \sqrt{\alpha} \, du.
\end{equation}
Inserting these into (\ref{e:meanphi2}) yields
\begin{align}
\label{e:uint}
\mean{\tphi(y)} & \approx
  q \kappa \sqrt{\alpha} \int_0^{\infty} \, du \, \exp
  \bigl[-\frac{\pi}{24} (\alpha u^2 + 2 R_y u)^2 \bigr] \\ & \approx
  \frac{\sqrt{6}}{2} \, q \kappa \, \sqrt{\alpha} \frac{1}{R_y} ,
\end{align}
since the integral is dominated by the region in which $u \ll R$,
where the integral is approximately Gaussian. Note that $R_y$ is
approximately $R$, because the amplitude of the source $a \ll R$ and
$d \ll R$.  Inserting this result into eqn.(\ref{e:meanF}) we obtain
our result for the mean output,
\begin{equation}
\label{e:meanFresult}
\mean{\tF} \approx \sqrt{\frac{1+\vel}{1-\vel}}  \frac{q}{4 \pi R} TA\,d\, \approx  (1+\vel) \frac{q}{4 \pi R} TAd,
\end{equation}
in the low velocity approximation, where we have set $\kappa=1 /( 2
\sqrt{6}\pi)$.  We see that the mean output in the discrete model
matches the continuum result (\ref{e:continuumoutput}) given above.
We have approximated the spatial distance from source points to
detector points as $R$ in all cases when finding $\tphi(y)$, but this
is basically the same approximation we made in the continuum case.

It is of interest to find the size of the correction to the continuum result.  The main extra approximation used in the discrete case was the approximation of the integral in equation (\ref{e:uint}) as Gaussian.  Instead of throwing away all but the Gaussian factor, we can include the next most significant factor:
\begin{equation}
\mean{\tphi(y)} \approx
  q \kappa \sqrt{\alpha} \int_0^{\infty} \, du \, \exp
  \bigl[-\frac{\pi}{6} ( R_y \alpha u^3 +  R_y^2 u^2 ) \bigr].
\end{equation}
Solving this integral by symbolic computation, and expanding in powers of $1/R$, shows a first term equal to the continuum result, and a second term proportionally smaller by a factor of order $1/R^2$.  This $R$ is a large distance expressed in Planck units, and so the extra approximation introduced in the discrete case is insignificant.
%Approximating the intersection of the light-cone of a source point with the detector as a plane, and some other approximations, were also made in the continuum case.  Experiments sensitive to such discrepancies are far beyond current sensitivities.

Just as in the continuum case, we cancel the DC component of the
response by adding a static, negative charge, modelled
on the causal set as a
path close to the origin. The mean of the
total response is the sum of the means of the responses
to the two sources and is
\begin{equation}
\label{e:meanFtotal}
\mean{\tF_{\textrm{total}}} \approx \vel  \frac{q}{4 \pi R} TA\,d\, .
\end{equation}

\subsection{Fluctuations in the detector output}

We have shown above how, in our model of propagation, the average
value (over sprinklings) of the detector signal does not significantly
differ from the signal in the standard case.  However, this result
must be strengthened.  We want to make sure that the result does not
differ significantly not just for the average, but for a typical
sprinkling (moreover we want to make sure that the model
predicts the correct result for many repeats of the experiment).  To
this end we calculate the variance of the signal.
We start by calculating the variance when the source consists
of a single oscillating positive charge.

As stated above, the detector output $\tF$ is proportional to the
number of links from the source to the detector.  If the existence of
any link was uncorrelated to the existence of any other link,
we would expect a variance of $q \kappa \mean{\tF}$
which is small. Correlations do exist, however, intuitively they will not
be large.  The result of this intuition -- that the fluctuations, as a
proportion of the signal, are small for reasonable values of the parameters -- is
correct, as we will now show.

Let us return to the expression (\ref {e:Frandvar}) for the output as
a random variable in the sprinkling process on $\Minkowski^4$.  We
have already calculated $\mean{\tF}$ on this basis.  There, the three
random variables in the expression for $\tF$ are uncorrelated.  Now we
seek to calculate $\var(\tF)= \mean{\tF^2} -\mean{\tF}^2$.  The term
$\mean{\tF^2}$ can be written

\begin{equation}
\mean{\tF^2} = q^2 \kappa^2 \int_P \int_{\cD} \int_P \int_{\cD \, \ab} \bra \sigma(ds_1) \chi(d^4y_1) \sigma(ds_2) \chi(d^4y_2)
\zeta(x(s_1),y_1) \zeta(x(s_2),y_2) \ket.
\end{equation}
where $s_1$ and $s_2$ are both proper time along $P$, $y_1$ and $y_2$
range over the detector $D$, and we have used the symbol $\ab$ as
shorthand for the conditions $x(s_1) \prec y_1$, $x(s_2) \prec y_2$.  In
this case, not all of the random variables are uncorrelated.  Firstly,
$\chi(d^4y_2)$ is correlated to $\zeta(x(s_1),y_1)$ as $y_2$ can fall
inside the causal interval $I(x(s_1),y_1)$.  In this case if a point is sprinkled
in $d^4y_2$, $\zeta(x(s_1),y_1)=0$.  Therefore there is
no contribution to the integral from the range in which $y_2 \in
I(x(s_1),y_1)$.  Using similar reasoning we can put 4 restrictions on the
region of integration without affecting its value:
\begin{align}
x(s_1) &\notin I_2, \label{e:x1}\\
 y_1 &\notin I_2, \label{e:y1}\\
x(s_2) &\notin I_1, \label{e:x2}\\
 y_2
&\notin I_1, \label{e:y2}
\end{align}
where we introduce the notation $I_1:=I(x(s_1),y_1)$ and
$I_2:=I(x(s_2),y_2)$.  We will use the symbol $\bb$ as shorthand for
these restrictions.  Inside the remaining region of integration, all $\chi$ and $\sigma$ variables
are uncorrelated with $\zeta$ variables.

Secondly, the $\chi$ and $\sigma$ variables can be correlated with
each other, as they do not all range over disjoint regions.  When
$y_1$ is in $d^4y_2$ there is a correlation.  Taking this correlation
into account is crucial to obtain the correct result (the situation is
similar to accounting for the self-correlations in a discrete set of
variables).  Given a sprinkled point at $y_1$, when $y_2 \neq y_1$,
the probability of finding a sprinkled point in $d^4y_2$ is
infinitesimal.  But when $y_1 = y_2$ the probability is 1, explaining
the importance of this correlation.  We deal with this by
splitting the integral into a sum of 4 terms:
\begin{equation}
\mean{\tF^2} = J_1 + J_2 + J_3 + J_4 \, .
\end{equation}
These terms take the form
\begin{equation}
\label{e:jterms}
J_i = q^2 \kappa^2 \int_P \int_D \int_P \int_{D \, \ab \, \bb \, \jib}
\bra \sigma(ds_1) \chi(d^4y_1) \sigma(ds_2) \chi(d^4y_2)
\zeta(x(s_1),y_1) \zeta(x(s_2),y_2) \ket \, ,
\end{equation}
with the additional $\bb$ bounds from
eqns. (\ref{e:x1}-\ref{e:y2}) above, and also some new
bounds $\jib$ for each term:
\begin{align}
\jab \, &: \, s_1=s_2, \, y_1=y_2    \label{e:ab1}   \\
\jbb \, &: \, s_1 \neq s_2, \, y_1=y_2  \\
\jcb \, &: \, s_1=s_2, \, y_1 \neq y_2 \label{e:ab3} \\
\jdb \, &: \, s_1 \neq s_2, \, y_1 \neq y_2 \, .
\end{align}
In each term, once the obvious identifications have been made there are
no remaining correlations between $\chi$ and $\sigma$ variables.
Finally there are correlations between the two $\zeta$ variables,
which will be dealt with in due course.  We will also define
\begin{equation}
K_4=J_4 - \mean{\tF}^2,
\end{equation}
so that
\begin{equation}
\label{e:varl}
\var(\tF) = J_1 + J_2 + J_3 + K_4.
\end{equation}
We now proceed to calculate these four terms.

Let us start with the $J_1$ term.  Due to the $\jab$ bound, which can
be used to eliminate $s_2$ and $y_2$, this term can be written
\begin{align}
 J_1 &= q^2 \kappa^2 \int_P \int_{\cD \, x \prec y} \bra \sigma(ds)
 \chi(d^4y) \zeta(x,y) \ket , \\ &= q \kappa \mean{\tF}.
\label{e:J1result}
\end{align}

Consider now the term $J_2$.  Calling $y_1=y_2=y$ we have
\begin{equation}
\label{e:j2}
J_2 = q^2 \kappa^2 \int_P \int_P \int_{\cD \, \ab \, \bb}
\bra \sigma(ds_1) \sigma(ds_2) \chi(d^4y)
\zeta(x(s_1),y) \zeta(x(s_2),y) \ket
\end{equation}
where the bounds should be interpreted with $y_1=y_2=y$.  These bounds
are incompatible. Bound $\ab$ implies that $x(s_1) \prec y$ and
$x(s_2) \prec y$.  But all points on $P$ are causally related,
including $x(s_1)$ and $x(s_1)$, so $x(s_2) \prec x(s_1) \prec y$ or
$x(s_1) \prec x(s_2) \prec y$.  This contradicts either (\ref{e:x1})
or (\ref{e:x2}) of the $(b)$ bounds.  We therefore have
\begin{equation}
J_2 = 0.
\end{equation}

The term $K_4$ suffers a similar fate:
\begin{align}
\label{e:j4}
J_4 &= q^2 \kappa^2 \int_P \int_{\cD} \int_P \int_{\cD \, \ab \, \bb}
\bra \sigma(ds_1) \chi(d^4y_1) \sigma(ds_2) \chi(d^4y_2)
\zeta(x(s_1),y_1) \zeta(x(s_2),y_2) \ket \\
    &=q^2 \kappa^2 \int_P \int_{\cD} \int_P \int_{\cD \, \ab \, \bb}
    ds_1 \, d^4y_1 \, ds_2 \, d^4y_2 \bra \zeta(x(s_1),y_1)
    \zeta(x(s_2),y_2) \ket \\ &= q^2 \kappa^2 \int_P \int_{\cD} \int_P
    \int_{\cD \, \ab \, \bb} ds_1 \, d^4y_1 \, ds_2 \, d^4y_2
    \exp(-|I_1 \cup I_2|),
\end{align}
where $I_i$ is the causal interval between $x(s_i)$ and $y_i$ as before,
 and $|.|$ indicates the volume of a
region.  The last line comes from seeing that the product
$\zeta(x(s_1),y_1) \zeta(x(s_2),y_2)$ is only 1 when both of these
intervals are empty of points, $\exp(-|I_1 \cup I_2|)$ being the
probability for this to happen.  Subtracting $\mean{\tF}^2$,
\begin{equation}
\mean{\tF}^2 = q^2 \kappa^2 \int_P \int_{\cD} \int_P \int_{\cD \, \ab}
    ds_1 \, d^4y_1 \, ds_2 \, d^4y_2 \exp(-|I_1|-|I_2|)
\end{equation}
gives
\begin{equation}
K_4 \leq q^2 \kappa^2 \int_P \int_{\cD} \int_P \int_{\cD \, \ab \,
\bb} ds_1 \, d^4y_1 \, ds_2 \, d^4y_2 \Bigl[\exp(-|I_1 \cup I_2|) -
\exp(-|I_1|-|I_2|) \Bigr],
\end{equation}
where we have added the $\bb$ bound to the $\mean{\tF}^2$ integral,
which can only increase the value of this upper bound on $K_4$. The
integrand here is zero unless $I_1 \cap I_2 \neq \emptyset$.  It can
be seen that there is no region inside the bounds satisfying this
condition.  As in the $J_2$ case, we use the assertion that $x(s_1)
\prec x(s_2)$ or $x(s_2) \prec x(s_1)$.  Let us assume $x(s_2) \prec
x(s_1)$, without loss of generality, as the conditions are symmetric
between $s_1$ and $s_2$.  The condition $I_1 \cap I_2 \neq \emptyset$
implies $x(s_1) \prec y_2$.  Together with $x(s_2) \prec x(s_1)$ this
gives $x(s_1) \in I_2$, contradicting (\ref{e:ab1}) in the $\ab$
bounds.  Hence,
\begin{equation}
K_4 \leq 0.
\end{equation}

The only remaining term is $J_3$:
\begin{equation}
\label{e:j3}
J_3 = q^2 \kappa^2 \int_P \int_{\cD} \int_{\cD \, \ab \, \bb} ds \,
d^4y_1 \, d^4y_2 \exp(-|I_1 \cup I_2|),
\end{equation}
where now, because of the identification $s_1=s_2=s$, we have $I_1=
J^+(x(s)) \cap J^-(y_1)$ and similarly for $I_2$.  Noting that $|I_1 \cup
I_2| \geq (|I_1| + |I_2|)/2$, we can write
\begin{equation}
\label{e:j3b}
J_3 \leq q^2 \kappa^2 \int_P \int_{\cD \, \ab} ds \, d^4y_1
\exp(-\frac{1}{2}|I_1|)
\int_{\cD \, \ab \, \bb} d^4y_2 \exp(-\frac{1}{2}|I_2|).
\end{equation}
Let us define $r_i=y^1_i - x^1(s)$ and $t_i=y^0_i - x^0(s)$, and
$u_i=t_i-r_i$ and $v_i=t_i+r_i$ (note that in this case we will be holding the $x$ variable constant while integrating over $y$, conversely to the method by which the mean was calculated).  As a step to
bounding $J_3$, consider
\begin{equation}
X(x(s),y_1) = \int_{\cD \, \ab \, \bb} d^4y_2
\exp(-\frac{1}{2}|I_2|).
\end{equation}
This range of integration for $y_2$ is much larger than necessary.
Let us assume that the light-cone of $x(s)$ does not intersect the
initial or final spatial boundaries of the detector region
(for the points in $P$ at which this does not hold, the region of $\cD$ integrated over is smaller, so this results in an upper bound on $X(x(s),y_1)$ for all points in $P$).  In this case, the bounds on the integral are
$R-x^1(s)<r_2<R-x^1(s)+d$, $0<y_2^2<\sqrt{A}$ and $0<y_2^3<\sqrt{A}$.
As before we will use $\tau(x(s),y_2)^4 \approx (u_2v_2)^2$. The bound $\ab$ gives $u_2>0$ in this case, and so we have
\begin{equation}
X(x(s),y_1) \lesssim \frac{A}{2} \int^{\infty}_{0} du_2
\int_{2(R-x^1)+u_2}^{2(R-x^1+d)+u_2} \, dv_2 \, \exp \bigl(-\frac{\pi}{48}(u_2
v_2)^2\bigr).
\end{equation}
As before we have used $\tau(s,y_2)^4 \approx (u_2v_2)^2$.
We have
\begin{align}
X(x(s),y_1) &\lesssim Ad \int^{\infty}_{0} du_2 \, \exp
\bigl(-\frac{\pi}{48}(u_2(2(R-x^1)+u_2))^2 \bigr) \\ &\lesssim %\approx
\sqrt{3} A d \frac{1}{R},
\end{align}
since $d$ and $x^1$ are negligible compared to $R$
and as in the calculation of (\ref{e:uint}) we have neglected
insignificant terms in the exponential.  Inserting this back into
(\ref{e:j3b}) we have
\begin{equation}
J_3 \lesssim \sqrt{3} q^2 \kappa^2 A d \frac{1}{R} \int_P \int_{\cD \,
\ab} ds \, d^4y_1 \exp(-\frac{1}{2}|I_1|).
\end{equation}
This integral is similar to
the one we have already calculated to find $\mean{\tF}$ in
eqns. (\ref{e:meanF}--\ref{e:meanFresult}).  We find
\begin{equation}
\label{e:J3result}
J_3 \lesssim  \sqrt{6}\,  q  \kappa A d \frac{1}{R} \mean{\tF} \, .
\end{equation}
Finally, summing up the remaining terms in eqns.(\ref{e:J1result}) and
(\ref{e:J3result}) to find $\var(\tF)$ from eqn.(\ref{e:varl}):
\begin{equation}
\label{e:var_postivepart}
\var(\tF) \leq  q \mean{\tF} (1 +  \sqrt{6} \kappa A d \frac{1}{R}).
\end{equation}

Now, $\var(\tF)$ represents the variance of the signal from only our oscillating positive charge.  We are not actually interested in $\var(\tF)$ but rather
in the variance of the {\textit{total}} detector response to
the oscillating positive charge and the static negative charge
together, $\var(\tF_{\textrm{total}}) = \var(\tF_+ + \tF_-)$,
where we have renamed $\tF$ above as $\tF_+$ as before.  This total variance is
\begin{equation}
 \var(\tF_{\textrm{total}})=\var(\tF_+) + \var(\tF_-) + 2 \, \Cov (\tF_+,\tF_-),
\end{equation}
where $\Cov$ is the covariance function,
\begin{equation}
\Cov (\tF_+,\tF_-) = \mean{(\tF_+ - \mean{\tF_+}) (\tF_- - \mean{\tF_-})} .
\end{equation}
It is a standard corollary of the Cauchy-Schwarz inequality
 that
\begin{equation}
\Cov (\tF_+,\tF_-)^2 \leq  \var(\tF_+) \var(\tF_-),
\end{equation}
and so we have
\begin{align}
\label{e:var_tot_from_cz}
 \var(\tF_{\textrm{total}}) &\leq \var(\tF_+) + \var(\tF_-) + 2 \sqrt{\var(\tF_+) \var(\tF_-)} \\
                                            &< 4 \var(\tF_+),
\end{align}
where the result follows from (\ref{e:var_postivepart}) and (\ref{e:meanFresult}) : $\var(\tF_-)$ is the same function as $\var(\tF_+)$ with the source velocity $\nu$ set to zero
so $\var(\tF_-) \approx \var(\tF_+)$ for low velocities.

The standard deviation of the total signal as a proportion of the output is
\begin{equation}
\frac{\sqrt{\var(\tF_{\textrm{total}})}}{\mean{\tF_{\textrm{total}}}}
\sim  \frac{\sqrt{4 \var(\tF_+)}}{\mean{\tF_{\textrm{total}}}}.
\end{equation}
Using our result (\ref{e:var_postivepart}) for $\var(\tF_+)$  (and the value set earlier for $\kappa$) gives
\begin{equation}
\label{e:relative_fluc_result}
\frac{\var(\tF_{\textrm{total}})}{\mean{\tF_{\textrm{total}}}^2}
\lesssim
16 \left[\frac{\pi R}{\vel^2 TAd} +  \frac{1}{2\vel^2 T}\right]\,.
\end{equation}

We can now input some values for the parameters to estimate the fluctuations for a
model source and a model detector.  We will obtain an upper bound on
the fluctuations by making unrealistically restrictive estimates in some cases.
 For a description of a relevant source,
we look to discussion of coherence of electromagnetic radiation from extragalactic sources in the literature.  One coherent source that has been cited \cite{Lieu:2003ee} is the Active Galactic Nucleus PKS 1413+135 \cite{Perlman:2002fi}. This source is at a distance of order one Gigaparsec, and coherence was detected in radiation of wavelength $1.6 \, \mu m$.  Therefore $R$ is set to be of order $10^{60}$ Planck lengths, and in order that the time resolution of detection be much better than the period of the radiation (one of the unrealistically restrictive estimates), we set $T$ to be $10^{27}$
Planck units.  The radiation from AGNs is thought to be synchrotronic, meaning that the source
electrons have velocities close to $1$. This doesn't fit with our low velocity
assumption but we choose $\vel ~ 0.1$ as a compromise value for
which our expansions are still approximately valid
\footnote{
To fully satisfy scepticism on this point, we would have to repeat our calculations without this approximation.  This would necessitate dropping the further assumption that the duration of the detection is much shorter than the period of oscillation of the source: when the maximum source velocity is close to the speed of light, the field from the postive charge would vary from much greater than its value for $\vel =0$ to much smaller, and so it is possible that the fluctuations as a proportion of the dectector output would vary greatly over one oscillation.  Such a large value for the relative fluctuations at only one time in the oscillation would not in practice be measureable, however; it is more realistic to consider a longer detection for the cited source in any case.  Considering this, there is strong reason to think that this more complicated treatment would not lead to a different conclusion.  When $\vel$ is positive, the ratio of fluctuations to detector output become even smaller as $\vel$ increases towards the relativistic regime, and when $vel \approx -1$ the contribution to the total detector output (and fluctuations) will be small.
}
.
As for the spatial size of the detector, $\sqrt{A}$ and $d$
can safely be set to be of order 1 nanometer,
or $10^{26}$ Planck units.  With these values of the parameters
the first term inside the brackets in equation
(\ref{e:relative_fluc_result}) is of order $10^{-43}$ and the second
is of order $10^{-25}$.
With these values the standard deviation of the total signal is about one part in $10^{12}$.
This order of fluctuation is well below that which is possible to detect even in more realistic set-ups than this nano-size and femto-duration detection.

This shows that, with these values for the parameters of our model, the fluctuations are a small part of the total signal.  Thus for a typical sprinkling, it is justified to claim that the results of the discrete model match the continuum model sufficiently well for the difference to be undetectable.  The conclusion is that the spatial and temporal
coherence of waves from distant sources is preserved in this model.

It is interesting to ask if models of this type can give good results for all types of source.  In this connection, it should be noted that certain approximations and assumptions, mostly unfavourable to the eventual outcome, have been made to achieve the result given above for the distant AGN.  The bound on fluctuations coming from equation (\ref{e:relative_fluc_result}) is well above that which could reasonably be achieved in such models, for three main reasons.  Firstly, the values for the quantities involved are all set to give a higher result for the fluctuations than is realistic, to give an upper bound; the only one that might conceivably be less favourable for other sources is $\vel$.  Secondly the bound on $\var(\tF_+)$ given in (\ref{e:var_postivepart}) is fairly crude: not only could a smaller upper bound be found for $J_3$, but the $K_4$
term could be less than zero.
Thirdly, making the model more realistic, by considering detection lasting for multiple oscillation periods, and/or spatially extended
sources\footnote{
An extended source could be built by adding many positive and negative particles, in which case a bound on the total variance similar to (\ref{e:var_tot_from_cz}) would still hold.  The resulting bound would be proportional the square of the number of particles, and  $\tF_{\textrm{total}}$ would of course be proportional to the number of particles, leaving (\ref{e:relative_fluc_result}) unchanged.  However, in this case the result analogous to (\ref{e:var_tot_from_cz}) would be a very much looser bound, as signals generated from significantly separated source particles would not be highly correlated.  The relative fluctuations in such a model would arguably be greatly reduced as a result.
}, would be expected to greatly reduce the relative size of
the fluctuations.  Thus it is reasonable to
conjecture  that models of this type could be constructed to model
many  realistic electromagnetic sources\footnote{
Indeed, the situation might reasonably be expected to be better for a similar discretisation of the electromagnetic Green's functions, if the calculations were otherwise similar to those above.  In the continuum, for the scalar field, the DC and AC parts of the signal have the same order of magnitude when $\vel \sim 1$, whereas in the electromagnetic case the the AC part will be larger by a factor of $R$.  This would suppress the total variation of the signal but not the total mean signal.
This could be analysed by investigating the causet version  of the continuum model
suggested at the end of section 2 in which the field $\phi$ is treated as an
electostatic potential.
}. In any case, the main aim here was to show that the model is consistent with
the coherence of waves from distant sources.

% Working for the bounds:
% T = 10^{-15} s * 10^{43} t_p s^{-1} = 10^{28}   , t_p is Planck time
% R = 10^9 Pc 10^{16} m Pc^{-1} * 10^{35} l_p m^{-1} = 10^{60}   , l_p is Planck length
% d= 10^{-9} m * 10^{35} = 10^26
% A = d^2 = 10^52
% \vel = 10^{-10} ?  Estimate for 10 MHz??? 10^-7 period so T=10^-9 makes sense
% R/(TAd) = 10^{60 - 28 - 26 - 52} = 10^{-46}
% 1/T = 10^{-28}
% take the square root and divide these by 10^{-2 }
% \sqrt{R/(TAd)} / \vel = 10^{-23+2} = 10^{-21}
% \sqrt{1/T} / \vel = 10^{-14+2} = 10^{-12}
\section{Conclusion}

We have reviewed a simple model of the propagation of a scalar field from source to detector in a Minkowski spacetime,
and demonstrated a model of the same physical
system in which the continuous spacetime is replaced with a causal set.
The main result is that the output of the detector is not significantly
affected by the introduction of discreteness.  This being so, the
phase of the source as it oscillates is faithfully recorded at all
astrophysical and cosmological distances: there is no
loss of (classical) coherence. Nor is there any change in
the dispersion relations. The discreteness \textit{will}
eventually show up when the distance, $R$, between source and detector
becomes large enough that the mean number of links from source to detector
region is of order one but the signature of the discreteness will be
a ``cutting out'' of
the signal altogether rather than any loss of coherence.
Unfortunately, for this to happen for any reasonably
sized detector, $R$ would have to be a
super-horizon distance and in any case the signal would by
then be undetectably tiny.

The relevance for quantum gravity phenomenology of a model premised on
a fixed background structure may be questioned. After all, the
continuum seems likely to result from a ``quantum superposition'' of
causal sets, whatever that turns out to mean.  However, the purpose of this
article was to show that discreteness {\textit {per se}} does not
pose any problem for the coherent propagation of waves over long
distances. Claims about ``generic'' phenomenological predictions
from quantum gravity will need to take account of this. For example,
it cannot now be claimed that
a fuzziness in spacetime metric relations necessarily disrupts Huygen's
principle nor that modified dispersion relations always result from
  the existence of a minimum length scale.
We emphasise, again, that the maintenance of coherence and
the usual dispersion relations hold for any fixed causal set
that arises with relatively high probability from the sprinkling process
and is not the result of any averaging over causets.

%\footnote{Although a total loss of coherence would be certainly be
%contrary to observation, this is not to say that it would be useless
%to consider models whose predictions differed more significantly from
%those of standard theory than the model given here.  For example, a
%Lorentz invariant ``dispersion'' in the energy of photons analogous to
%that proposed for massive particles in \cite{Dowker:2003hb} may well
%give other interesting phenomenology, and still be small enough to
%agree with observations of coherence of light from extragalactic
%sources.  This is especially true considering the classical coherence
%preserving effects noted in \cite{Coule:2003td}.}. It is interesting to see an explicit case in which some fuzziness in spacetime metric relations can be introduced without affecting coherence.  This shows that the effect is not a generic feature of such scenarios.
%This is not to suggest that totally generic predictions are the only ones of use;
%as in the case of the discreteness of matter, it seems to be a sensible approach to make and test many different models that embody different physical hypotheses.

The crucial aspect of causal set discreteness that
distinguishes it from other approaches is that it respects Lorentz
invariance.
This is the key to preserving the usual dispersion relation in the above
model: even without a detailed model, we would have claimed that
variance in the speed of light as in \cite{Lieu:2003ee}
could not be present in a causal set model unless
it were introduced ``by hand'' somehow.
The concrete model is particularly illuminating because it
shows explicitly  that the speed of
photons from distant sources is not affected by causet
discreteness due to the existence in the causet of
structure -- the links -- which traces lightcones extremely accurately.
It is hoped that the study of Gamma ray pulses from
Blazars and Gamma ray bursts \cite{AmelinoCamelia:2008qg,Amelino-Camelia:2004yd} will either detect, or place
stringent limits on, variations in $c$ and it remains to be seen which
discrete models of propagation will be consistent with these observations.

We have thus seen that, due to the tininess of the Planck scale and
the Lorentz invariance of the approach, causal set models can give
results that are experimentally indistinguishable from the continuum.
This is simultaneously
encouraging and disappointing. Encouraging because every such result
provides more evidence for the ``Hauptvermutung'' (Central
Conjecture) of causal set theory, that causal sets can indeed
be the deep structure of Lorentzian spacetimes.
The disappointment
arises because providing experimental evidence for causal set
theory will require predictions that are
at variance with those made in the continuum.
Though the simple model presented here seems not
to be able to produce such phenomenology, other models
of matter propagating on causal sets do hold out the hope of producing
observable signatures of discreteness. One promising class of
models is based on the idea that if spacetime is a causal set then
this should cause small, random, Lorentz invariant
fluctuations in the motion of
particles through spacetime. For massive particles this results in
a diffusion in momentum \cite{Dowker:2003hb, Philpott:2008vd}
and for massless particles, both
a diffusion and drift in frequency \cite{Philpott:2008vd}.
These effects are potentially observable and astrophysical and cosmological
data has been used to bound the parameters of the models
\cite{Dowker:2003hb,Kaloper:2006pj, Mattingly:2007be, Philpott:2008vd}.
Beyond these models, the discovery of a discrete, Lorentz invariant,
D'Alembertian operator \cite{Henson:2006kf,Sorkin:2007qi}
means that a discrete scalar
wave equation can be solved and the behaviour of
waves and wave packets investigated, providing more realistic models
of matter and further opportunities for
predicting phenomena that could reveal a fundamental spacetime discreteness.

The authors are grateful to Ted Jacobson for a helpful remark,
and Tom Kibble for discussions that motivated
and informed this research.
RDS was partly supported by NSF Grant PHY-0404646.
Research at Perimeter Institute for Theoretical Physics is supported in
part by the Government of Canada through NSERC and by the Province of
Ontario through MRI.
FD was partly supported by Marie Curie Research and Training Network
``Random Geometry and Random Matrices:
From Quantum Gravity to Econophysics'' (MRTN-CT-2004-005616)
and Royal Society Grant IJP 2006/R2. FD is grateful to the
Perimeter Institute for Theoretical Physics for hospitality during
the writing of this paper.

\bibliographystyle{../Bibliography/JHEP}
\bibliography{../Bibliography/refs}

\end{document}